\documentclass{article}

\usepackage{natbib}

\usepackage[preprint]{neurips_2025}

\usepackage[utf8]{inputenc} % allow utf-8 input
\usepackage[T1]{fontenc}    % use 8-bit T1 fonts
\usepackage{hyperref}       % hyperlinks
\usepackage{url}            % simple URL typesetting
\usepackage{booktabs}       % professional-quality tables
\usepackage{amsfonts}       % blackboard math symbols
\usepackage{nicefrac}       % compact symbols for 1/2, etc.
\usepackage{microtype}      % microtypography
\usepackage{xcolor}         % colors
\usepackage{graphicx}
\usepackage{amsmath,amsthm,amssymb}
\usepackage{algorithm}
\usepackage{algorithmic}

\usepackage{listings}

\definecolor{lightgray}{rgb}{0.95,0.95,0.95}

\title{Ricci Matrix Comparison for Graph Alignment: A DMC Variation}

\author{%
  Ashley M.~Wang\thanks{Email address after June, 2025: ashmwang@stanford.edu.
    }\\
  Department of Mathematics\\
  Dartmouth College\\
  Hanover, NH 03755 \\
  \texttt{ashley.m.wang.25@dartmouth.edu} \\
  \And
  Peter Chin \\
  Thayer School of Engineering \\
  Dartmouth College \\
  Hanover, NH 03755 \\
  \texttt{peter.chin@dartmouth.edu} \\
}

\begin{document}

\maketitle

\begin{abstract}
  The graph alignment problem explores the concept of node correspondence and its optimality. In this paper, we focus on purely geometric graph alignment methods, namely our newly proposed Ricci Matrix Comparison (RMC) and its original form, Degree Matrix Comparison (DMC). To formulate a Ricci-curvature-based graph alignment situation, we start with discussing different ideas of constructing one of the most typical and important topological objects, the torus, and then move on to introducing the RMC based on DMC with theoretical motivations. Lastly, we will present to the reader experimental results on a torus and a complex protein-protein interaction network that indicate the potential of applying a differential-geometric view to graph alignment. Results show that a direct variation of DMC using Ricci curvature can help with identifying holes in tori and aligning line graphs of a complex network at 80--90+\% accuracy. This paper contributes a new perspective to the field of graph alignment and partially shows the validity of the previous DMC method.
\end{abstract}

\section{Introduction}
Graphs are fundamental geometric objects. Mathematically, graph theory is a subfield of combinatorics. In computer science, graphs (or networks) form an important type of data structure to store information on connections. They have been used in social and biological studies prevalently since connections are inherently present. Our paper concerns the graph alignment problem, which assigns a correspondence between nodes on two networks. We assume upfront that there is common structure that is suitable for alignment. The most ideal solution would be assign all nodes that are supposed to be identical accurately. For example, if we had two relational networks of human tissues and were to align them, the optimal solution would be to correspond all tissue groups correctly. 

Graph alignment is useful many ways, with applications across various domains. For example, graph alignment can identify the same user across multiple social networks, providing insights into user behavior and preferences; it is also commonly used for community detection. In biology, it helps with studying protein-protein interaction (PPI) networks \citep{Bayati2013}. Due to their complexity and heterogeneity, PPI networks are often used to test graph alignment algorithms \citep{Wang2025}. Graph alignment is also highly related to the subgraph isomorphism problem in mathematics and computer science, which is an NP-complete problem \citep{Wang2025}. In this paper, we propose the possibility of using graph curvature as a tool for graph alignment. Namely, we propose Ricci Matrix Comparison (RMC), which is a method inspired by Degree Matrix Comparison (DMC) \citep{Wang2025}. The DMC method uses a matrix to record degrees of neighbors for each node on a graph, and has been shown to be effective for aligning heterogeneous graphs. Our proposed method, RMC, replaces these degrees with the Forman-Ricci curvatures, which will be introduced in Section~\ref{sec:curvatureintro}, of neighboring nodes. This method is intended to be for graphs that are suitable as models of discretized smooth surfaces. Both DMC and RMC will be introduced with detail in Section~\ref{sec:DRMC}.

We start with simple geometric objects. We apply RMC to a pair of tessellated tori through regular tiling for experiment. We devoted Section~\ref{sec:tessellation} to discussing tessellations in both a plane and three-dimensional space, and these tilings of a one-hole torus can easily be generalized to higher-genus tori. We focus on regular tessellations, which allows us to build a discretized torus using basic units.

Afterwards, we develop theoretical statements in Section~\ref{sec:theory} that motivate the use of curvature for graph alignment. Along with that, we apply RMC to a discretized torus and show that it is at least effective for categorizing nodes on the same geometric object by location, which will also help us identify holes. Since the torus we construct are inherently standard and hence symmetric, we do not seek node ID alignment, but rather only focus on geometric identities. Additionally, we include a RMC alignment result on the line graph (see Section~\ref{sec:linegraph}) of a PPI network. Due to the local completeness of line graphs, their structures are closer approximations of continuous solid volumes (and hence surfaces). Our method demonstrates high performance in terms of accuracy, but this attempt has limitations at this stage because line graph generation is computationally expensive. We will present more details in Section~\ref{sec:results}. Understanding efficient ways to create a line graph could lead to another research project. 

\section{Forman-Ricci Curvature}
\label{sec:curvatureintro}
The word ``curvature'' is self-explanatory, generally speaking. However, the rigorous mathematical definitions can be lengthy and they vary. Additionally, curvatures are typically mentioned in the context of smooth surfaces. We introduce graph curvatures as discrete analogues of curvatures for smooth surfaces. The most common types of graph curvatures are the Forman-Ricci curvature and Ollivier-Ricci curvature. The Ollivier-Ricci curvature has been used in \citet{Sia2019} for community detection in complex networks, which shows that curvature can be utilized as a tool for alignment. However, we choose to use the Forman-Ricci curvature for a first attempt to construct a graph alignment algorithm that is both curvature-based and matrix-based because it is computationally cheap, even though it sacrifices some accuracy. 

The Forman-Ricci curvature is initially defined on edges. We assign a value to each of the edges which we call curvature (discretized Ricci curvature). For a weighted graph, the formula is 
\label{def:formanCurvature}
\begin{equation}
\textbf{Ric}(e) = w_{e}\left(\dfrac{w_{v_{1}}}{w_{e}} + \dfrac{w_{v_{2}}}{w_{e}} - \sum_{e_{l}\sim v_{1}}\dfrac{w_{v_{1}}}{\sqrt{w_{e}w_{e_{l}}}} - \sum_{e_{l}\sim v_{2}}\dfrac{w_{v_{2}}}{\sqrt{w_{e}w_{e_{l}}}}\right),
\end{equation}
which is taken from \citet{Weber2017}. $e$ is an edge between nodes $v_{1}$ and $v_{2}$, and $w_{e}, w_{v_{1}}, w_{v_{2}}, w_{e_{l}}$ are weights on edges and nodes. $e_{l}\sim v_{i}$ means the edge $e_{l}$ is adjacent to the node $v_{i}$. It follows that for unweighted graphs, the curvature for each edge is
\begin{equation}
\textbf{Ric}(e) = 2 - \text{deg}(v_{1}) - \text{deg}(v_{2})
\label{eqtn:unweightedCurvature}
\end{equation}
if we take all weights to be uniformly 1. According to \citet{Weber2017}, we can define a node version of curvature simply by taking the sum of all edge curvatures of edges attached to the node, or:
\begin{equation}
\textbf{Ric}(v) = \sum_{e_{l}\sim v}\textbf{Ric}(e_{l}).
\label{eqtn:formanNode}
\end{equation}

\section{Tiling Methods to Construct a Torus}
\label{sec:tessellation}
We will highlight several attempts to tile both a two-dimensional ring and a three-dimensional torus uniformly with regular shapes. Our goal is to have the ability to approximate the standard one-hole torus with regular tiling methods, and this can be generalized to higher-genus objects through simply connecting one-hole components. A regular tiling of a smooth torus with polygons is not possible, and approximations with polygons are important because this will be helpful in extracting graphs from the tiling. Our focus is not on finding new tiling methods, but rather to make clear the type of tiling we need to make the extracted graph later on to be useful. One could look into \citet{Shephard1977,Nagy2023} and Penrose tiling, for example, to appreciate the many types of geometric tiling, which is a fascinating topic that has been studied for a long time.

\subsection{Two-Dimensional Tiling}
\label{sec:2DTiling}
To examine two-dimensional tiling for an approximated ring, we point out that there are only three edge-to-edge regular tiling in a plane (tile with standard polygons only and of the same type), and the shapes that can do this are: equilateral triangles, squares, and regular hexagons. This type of tiling can also be called regular tessellation. 
\begin{figure}
  \centering
  \includegraphics[width=0.15\textwidth]{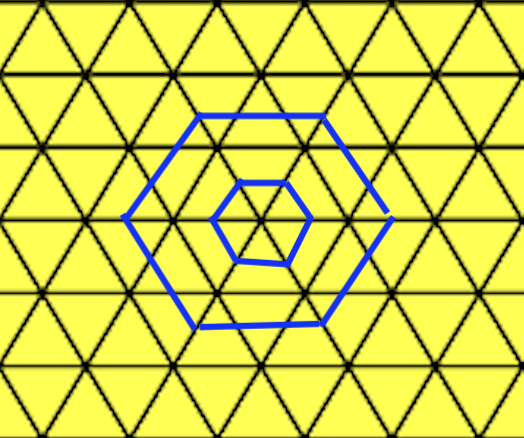}
  \caption{Equilateral triangular tiling. Blue hexagons indicate triangles that form a hexagonal ring.}
  \label{fig:triangtess}
\end{figure}
\begin{figure}
  \centering
  \includegraphics[width=0.15\textwidth]{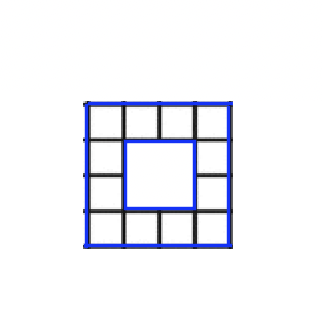}
  \caption{Square tiling for an approximate ring.}
  \label{fig:squaretess}
\end{figure}
We discuss the monohedral tilings using equilateral triangles and squares, and also mention dihedral tiling with the two shapes. While hexagons embody very interesting mathematical properties, we will not dive deep into it since it is not as basic a unit as a triangle or a square. An equilateral triangle tiling is given in Figure~\ref{fig:triangtess}. Possible two-dimensional rings are formed by triangles traced by the blue hexagons. This case is simple to implement and is a good basic model to start with. As we explore a curvature-based graph alignment method, this type of tiling gives us different curvatures for edges and nodes at different locations of the torus. Another simple approach is to use squares to build a square frame (see Figure~\ref{fig:squaretess}), which is topologically a one-hole object. However, the triangle approach fits our needs better, because for squares, being a node near the hole or farther away does not necessarily change the curvature.

As mentioned above, we can also create a mixed tiling of squares and triangles. This can be achieved through tiling six squares that each share one edge with a hexagon, then fill the gaps in the ring with equilateral triangles. Again, this tiling is valuable because we have variations in degrees of nodes (and hence the curvatures).  

\subsection{Three-Dimensional Tiling}
\begin{figure}
  \centering
  \includegraphics[width=0.2\textwidth]{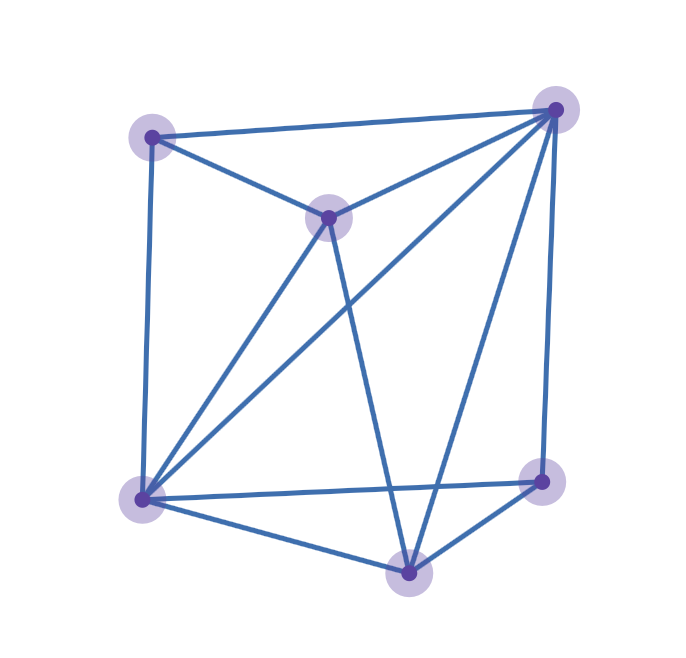}
  \caption{Triangulating a prism within three-dimensional torus.}
  \label{fig:prismTriang}
\end{figure}

We first create a two-dimensional ring, then lift a parallel layer and connect all corresponding nodes in the two copies. This already produces a torus-like shape. If one were looking for further triangulation, one could do so easily in the triangular case, where we obtain prisms that replace the original flat triangles in the plane. Within these prisms, we can add three edges to divide them into tetrahedrons (see Figure~\ref{fig:prismTriang}).

Essentially, we are looking for the simplest tiling method that produces a non-uniform degree (curvature) distribution over different locations of the torus. For example, we would want the curvature of a node near the hole to be different from that of a node on the outer brink.

\section{Ricci Matrix Comparison}
\label{sec:DRMC}
\subsection{Degree Matrix Comparison}
This method is introduced in \citet{Wang2025}. For the reader's convenience, we re-introduce here the major concepts needed to understand the method. We consider a pair of graphs with the same number of nodes. 

\paragraph{Definition 4.1} \emph{Degree Matrix:} Examining all node degrees on both graphs, we find the common maximum degree, which we denote as $m$. The ``degree matrix'' for each graph in the DMC is a $N \times m$ matrix, where $N$ is the total number of nodes on each graph. In each row of this matrix, we have a node's neighbors' degrees recorded in ascending order, left aligned and zero-padded to the right of all non-zero degree entries. 

After creating a degree matrix for both graphs, the authors in \citet{Wang2025} used the Hungarian algorithm to find an assignment between the two sets of all row vectors in the degree matrices to minimize total cost of alignment. The Hungarian algorithm is well known and also presented below. This is taken from \citet{Wang2025}.
\begin{algorithm}[H]
\caption{Hungarian Algorithm}
\label{algorithm:Hungarian}
\begin{algorithmic}[0]
\STATE \textbf{Input:} A cost matrix $C$ of size $n \times n$.

\STATE \textbf{Step 1: Subtract the row minimum.}
\FOR{each row $i$ in $C$}
    \STATE Subtract the minimum value of row $i$ from all elements in row $i$.
\ENDFOR

\STATE \textbf{Step 2: Subtract the column minimum.}
\FOR{each column $j$ in $C$}
    \STATE Subtract the minimum value of column $j$ from all elements in column $j$.
\ENDFOR

\STATE \textbf{Step 3: Cover all zeros with a minimum number of lines.}
\WHILE{not all zeros are covered}
    \STATE Identify rows and columns containing uncovered zeros.
    \STATE Cover all zeros using the minimum number of horizontal and vertical lines.
\ENDWHILE

\STATE \textbf{Step 4: Check the number of lines.}
\IF{the number of lines is equal to $n$}
    \STATE An optimal assignment is possible.
    \STATE Go to Step 6.
\ELSE
    \STATE Go to Step 5.
\ENDIF

\STATE \textbf{Step 5: Adjust the matrix.}
\STATE Find the smallest uncovered value.
\STATE Subtract this value from all uncovered elements.
\STATE Add this value to elements at intersections of covering lines.
\STATE Return to Step 3.

\STATE \textbf{Step 6: Make the assignments.}
\STATE Select zeros such that no two are in the same row or column.
\STATE Return the assignment and compute the total cost.

\STATE \textbf{Output:} An optimal assignment and the total cost.
\end{algorithmic}
\end{algorithm}
For a more intuitive view into the DMC process, one could consult Section 3.1 in \citet{Wang2025} for a concrete example. There is also an example in the same section on the construction of the degree matrices needed for input into the Hungarian algorithm. 

\subsection{Ricci Matrix Comparison}
As hinted earlier, we propose a possible variation to the DMC method. The method is simple: we replace all degrees in degree matrices introduced in the previous section with Forman-Ricci curvatures (see Equation \ref{eqtn:formanNode}). Instead of having degrees of a node's neighbors in each row, we replace these values with the Forman-Ricci curvatures of these same neighbors. We reorder these curvatures within each row to ascending order and pad zeros to the right again. Then we apply the Hungarian algorithm as previous researchers have done.

\begin{figure}
\centering
\includegraphics[width=0.2\textwidth]{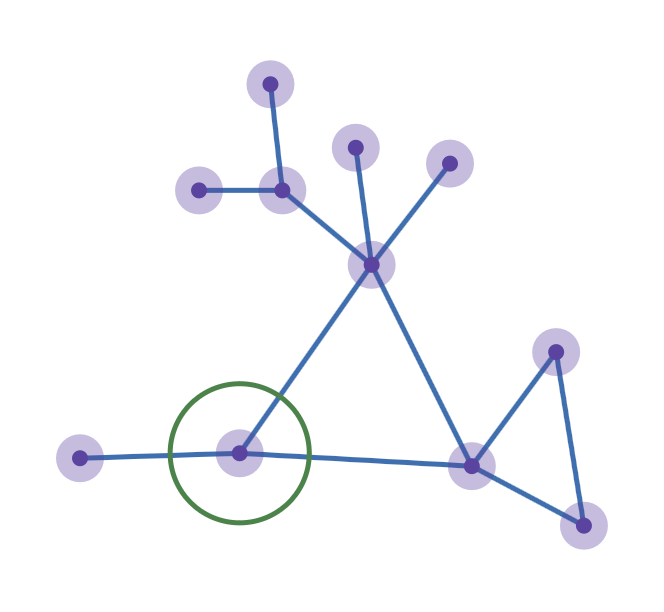}
  \caption{Example graph for demonstrating Ricci row vector construction.}
  \label{fig:RicciRowExample}
\end{figure}

\paragraph{Example 4.1} Suppose we try to look for the row vector in a Ricci matrix (analogue of degree matrix) for the circled node in Figure \ref{fig:RicciRowExample}. We first observe that the neighbors of this origin $O$ have degrees of 1, 4, and 5. Let us call these nodes $A$, $B$, and $C$. We try to calculate the Forman-Ricci curvature at each of these neighbors. For node $A$, the curvature is $2 - \text{deg}(A) - \text{deg}(O) = -2$ and the summation sign is dropped since it only has degree one. For nodes $B$ and $C$, we calculate their curvatures in the same way, summing up the curvatures on all adjacent edges. For node $B$, its adjacent edges have curvatures -5, -4, -4, and -7. Therefore, the Forman-Ricci curvature of node $B$ is their sum, -20. Following a same procedure, we get that the node curvature for $C$ is -27. So in a row vector, we have [-27, -20, -2, 0, 0, ..., 0]. The number of zeros we pad is dependent on the magnitude of $m$, the maximum degree on both graphs. The dimensions of the matrices remain the same as degree matrices since the number of nonzero entries of a row is still reflective of the degrees of the nodes. Note here that we will not encounter a complex connected graph with a node with curvature 0, since each node has at least a degree of 1.

\section{Theoretical Motivation}
\label{sec:theory}
In this section, we record an equation that we discovered for unweighted graphs that relates the Forman-Ricci curvature, graph Laplacian, and a slight variation of signature vectors in the degree matrix. We introduce these concepts one by one before deriving what we call the ``Curvature-Laplacian'' equation in Section~\ref{sec:dlFormula}. 

\subsection{Graph Laplacian}
\begin{figure}[h]
    \centering    \includegraphics[width=0.2\textwidth]{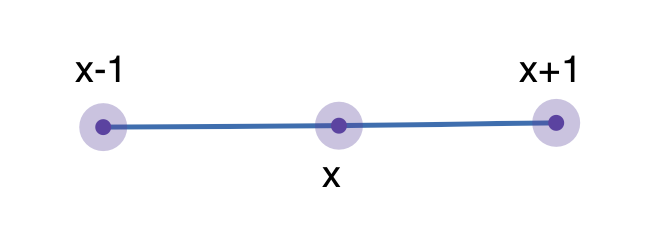}
    \caption{Illustration of one-dimensional graph.}
    \label{fig:oneDillustration}
\end{figure}

We introduce the graph Laplacian by deriving it as a discrete analog of the continuous Laplacian. We will first derive the discrete Laplace operator from the continuous Laplacian, then obtain the graph Laplacian from the discrete Laplace operator. A continuous Laplacian on function $f: \mathbb{R}^{n} \rightarrow \mathbb{R}$, can be written as
\begin{equation}
\Delta f = \sum_{i=1}^{n}\dfrac{\partial^{2}f}{\partial x_{i}^{2}}
\end{equation}
if $x_{i}$'s are Cartesian coordinates. In order to move from the continuous end to a discrete graph version, we incorporate numerical approximations. Using the Finite Difference Method, we can define the first derivative of $f$ with respect to $\mathbf x = (x_{1}, ..., x_{n})$ as
\begin{equation}
f'(\mathbf x) = \lim_{\mathbf{\epsilon} \rightarrow 0} \dfrac{f(\mathbf x + {\epsilon}) - f(\mathbf x)}{\epsilon}
\label{eqtn:Euler}
\end{equation}
where $\epsilon$ is a small $n$-dimensional perturbation vector. With Equation~\ref{eqtn:Euler}, we can derive a Laplacian with the central approximation in a one-dimensional setting:
\begin{equation}
\Delta f = f''(\mathbf{x}) = \lim_{\epsilon\rightarrow 0}\dfrac{\frac{f(\mathbf{x} + \epsilon) - f(\mathbf{x})}{\epsilon} - \frac{f(\mathbf{x}) - f(\mathbf{x} - \epsilon)}{\epsilon}}{\epsilon} = \lim_{\epsilon\rightarrow 0} \dfrac{f(\mathbf x + \epsilon) + f(\mathbf x - \epsilon) - 2f(\mathbf x)}{\epsilon^{2}}.
\label{eqtn:finiteDifference}
\end{equation}
In the discrete case for graphs, we take $\epsilon = 1$ (assuming edge lengths are uniformly 1) and let $\mathbf x$ be a node. Since this is a one-dimensional setting, $\mathbf{x} + \epsilon = \mathbf{x} + 1$ is the node to the right of $\mathbf x$ and similar definition holds for $\mathbf x - 1$. This is illustrated in Figure~\ref{fig:oneDillustration}.
Therefore, Equation~\ref{eqtn:finiteDifference} can be reformed as 
\begin{equation}
-\Delta f = [f(\mathbf x) - f(\mathbf x+\epsilon)] + [f(\mathbf x) - f(\mathbf x-\epsilon)].
\end{equation}

\begin{figure}[h]
    \centering    \includegraphics[width=0.15\textwidth]{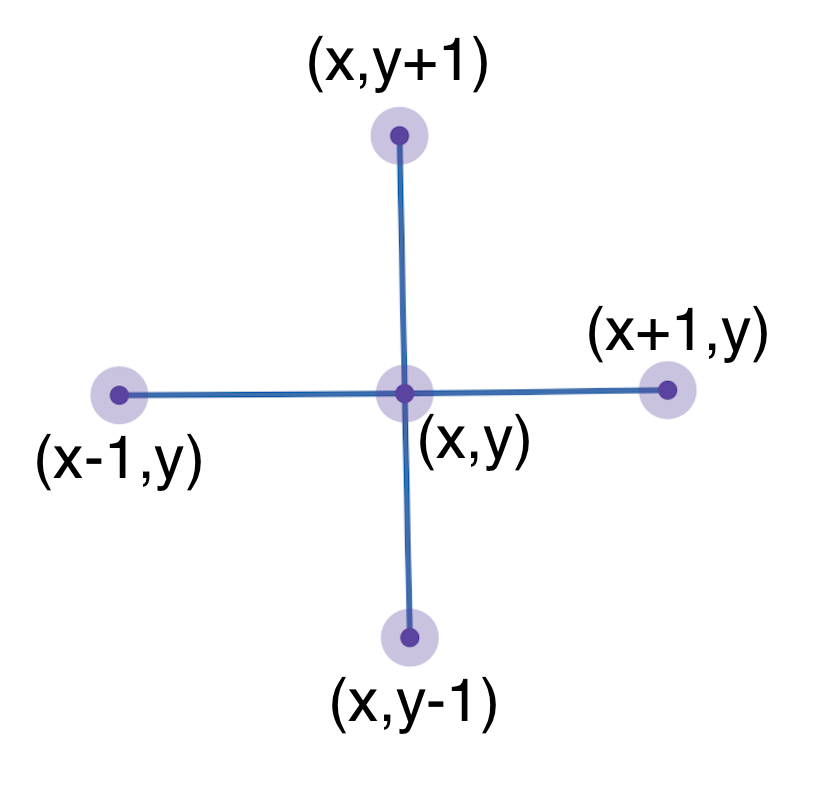}
    \caption{Illustration of two-dimensional graph.}
    \label{fig:twoDillustration}
\end{figure}

Suppose we extend this to a two-dimensional setting. Then 
\begin{equation}
\Delta f (x, y) =\dfrac{\partial^{2}f}{\partial x^{2}} + \dfrac{\partial^{2}f}{\partial y^{2}}, 
\end{equation}
which is, by slight variation of the one-dimensional case in Equation~\ref{eqtn:finiteDifference},  
\begin{equation}
\Delta f(x,y) = \lim_{\epsilon\rightarrow 0} \frac{f(x+\epsilon, y) + f(x -\epsilon, y) - 2f(x,y)}{\epsilon^{2}}+ \frac{f(x, y+\epsilon) + f(x, y -\epsilon) - 2f(x,y)}{\epsilon^{2}},
\end{equation}
which can be reduced to 
\begin{equation}
\Delta f(x,y) = \lim_{\epsilon\rightarrow 0} \dfrac{f(x+\epsilon, y) + f(x -\epsilon, y) + f(x, y+\epsilon) + f(x, y -\epsilon) - 4f(x, y)}{\epsilon^{2}}.
\end{equation}
Using again the idea that $\epsilon = 1$ and $(x,y)$ coordinates are nodes, we get that 
\begin{equation}
-\Delta f(x,y) = \sum_{(x_{l},y_{l})\in V}f(x, y) - f(x_{l}, y_{l})
\end{equation}
where the node set $V = \{(a,b)|(a,b) = (x,y) \pm (1,0) \text{ or } (x,y)\pm(0,1)\}$. The two-dimensional graph is illustrated in Figure~\ref{fig:twoDillustration}.

Following the idea above for the one-dimensional and two-dimensional cases, we can write 
\begin{equation}
-\Delta f (v) = \sum_{v_{i}\sim v}f(v) - f(v_{i})
\label{eqtn:discreteL}
\end{equation}
in general where $v_{i}\sim v$ denotes the nodes incident to $v$. 
% Conventionally, we define the discrete Laplace operator derived in Equation~\ref{eqtn:discreteL} without the negative sign, i.e.
% \begin{equation}
% \Delta f (v) = \sum_{v_{i}\sim v}f(v) - f(v_{i}).
% \end{equation}
By deriving a discrete version of the classic continuous Laplace operator, we have a basis to move on to obtain the graph Laplacian. Since $f: \mathbb{R}^{n} \rightarrow \mathbb{R}$ produces a real number, the vector $\mathbf f = [f(v_{1}), f(v_{2}),...,f(v_{n})]^{T}$ is an $n$-dimensional column vector with respect to a graph $G = (V, E)$ with the node set $V = \{v_{i}\}_{i=1}^{n}$. Then 
\begin{equation}
\Delta \mathbf f = [\Delta f(v_{1}), \Delta f(v_{2}), ..., \Delta f(v_{n})]^{T}
\end{equation}
and examining a single row (one entry) we have that
\begin{equation}
-\Delta f(v_{i}) = \sum_{v_{l}\sim v_{i}}f(v_{i}) - f(v_{l}) = \text{deg}(v_{i})f(v_{i}) - \sum_{v_{l}\sim v_{i}} f(v_{l}).
\label{eqtn:criticalStep}
\end{equation}
Let us define an $n \times n$ matrix (which we will later call the graph Laplacian)
\begin{equation}
L = D - A
\label{eqtn:graphL}
\end{equation}
where $D$ is a diagonal matrix with $\text{deg}(v_{i})$ as the only non-zero entry in the $i$-th row, and $A$ is the adjacency matrix (an $n \times n$ matrix where $a_{ij} = 1$ if edge $\{i,j\} \in E$ and $a_{ij} = 0$ otherwise). Then it is no coincidence that
\begin{equation}
L\mathbf f = -\Delta \mathbf f
\end{equation}
by Equation~\ref{eqtn:criticalStep}. As hinted above, $L$ in Equation~\ref{eqtn:graphL} is the \emph{graph Laplacian} (let us assume we are examining finite graphs). This definition of the graph Laplacian aligns with previous work \citep{Kostenko2021}.

\subsection{Variation of Signature Row Vector from a Degree Matrix}
In a DMC setting, we record degrees of neighbors of a node in ascending order in each row. Each of these row vectors can be viewed as a signature vector for a node. In this section, we propose a possible variation of this signature vector, allowing node labeling. In particular, we propose that the signature vector for a node $v_{i}$ can be modified to an $1 \times n$ row vector $\mathbf s$ such that 
\begin{equation}
s_{l}(v_{i}) =
\begin{cases}
\text{deg}(v_{l}) & \text{if} \{v_{l},v_{i}\} \in E,\\
\text{deg}(v_{i}) & \text{otherwise.} 
\end{cases}
\end{equation}
Let us call $\mathbf s$ the \emph{labeled signature vector} of $v_{i}$.

\subsection{The ``Curvature-Laplacian'' Equation}
\label{sec:dlFormula}
Finally, we have come to the section where we can introduce an interesting equation. The fact that the Ricci curvature at a node can be represented with the graph Laplacian (which is fixed for the same graph), the signature vector, and the degree of the node itself, motivates us to derive a curvature-directed graph alignment algorithm similar to the DMC.

\paragraph{Lemma
5.1} \emph{Curvature-Laplacian Equation:} Let $\text{Ric}(v)$ be defined as in Equation~\ref{eqtn:formanNode}. Let $L$ be the graph Laplacian. Let $\mathbf s$ be the labeled signature vector of a node $v_{i}$, as introduced in the previous section. Then 
\begin{equation}
\text{Ric}(v_{i}) - (L\mathbf s^{T})_{i} = 2\text{deg}(v_{i})(1 - \text{deg}(v_{i}))
\label{eqtn:CLE}
\end{equation}
is true for an unweighted graph. $(L\mathbf s^{T})_{i}$ is the $i$-th entry of the resulting vector. 

\begin{proof} By Equations~\ref{eqtn:unweightedCurvature} and \ref{eqtn:formanNode}, we know 
\begin{equation}
\text{Ric}(v_{i}) = \sum_{v_{l}\sim v_{i}}2-\text{deg}(v_{i}) - \text{deg}(v_{l}).
\end{equation}
Since $Lf = -\Delta f$, we have that
\begin{equation}
L\mathbf s^{T} = -\Delta \mathbf s^{T}  
\end{equation}
and hence 
\begin{equation}
(L\mathbf s^{T})_{i} = -(\Delta \mathbf s^{T})_{i} = \sum_{v_{l}\sim v_{i}}s_{i} - s_{l} = \sum_{v_{l}\sim v_{i}} \text{deg}(v_{i}) - \text{deg}(v_{l}).
\end{equation}
It follows naturally that 
\begin{equation}
\text{Ric}(v_{i}) - (L\mathbf s^{T})_{i} = 2\text{deg}(v_{i}) - \text{deg}(v_{i})^{2} - \sum_{v_{l}\sim v_{i}}\text{deg}(v_{l}) - \text{deg}(v_{i})^{2} + \sum_{v_{l}\sim v_{i}}\text{deg}(v_{l}) 
\end{equation}
which reduces to
\begin{equation}
\text{Ric}(v_{i}) - (L\mathbf s^{T})_{i} = 2\text{deg}(v_{i})(1 - \text{deg}(v_{i})).
\end{equation}
This completes the proof.
\end{proof}

\section{Line Graphs}
\label{sec:linegraph}
\paragraph{Definition 6.1} \emph{Line Graph:} A line graph $L(G)$ of an undirected graph $G = (V, E)$ is a graph that takes the edge set $E$ of $G$ to be the new node set so that $L(G) = (E, E')$, where $E'$ is the set of all connections between new vertices that were originally adjacent edges in $G$. 

Every local node, along with the edges connected to it, in $G$ becomes a locally complete component in the new line graph $L(G)$. This clustering helps form more compact local structures that will help with approximating a smooth surface with filled volume. We can use the line graph of graphs to help with graph alignment mainly due to the following result, proved in \citet{Whitney1932}. It essentially states that, in most cases, we can correspond each connected graph with one unique line graph.

\paragraph{Theorem 6.1} \emph{Whitney Isomorphism Theorem:} If the line graphs of two connected graphs are isomorphic, then the underlying graphs are isomorphic, except in the case of the triangle graph $K_{3}$ and the claw $K_{1,3}$, which have isomorphic line graphs but are not themselves isomorphic. 

$K_{n}$ is a complete graph with $n$ nodes. So $K_{3}$ is just a triangle. $K_{m,n}$ is the notation for a complete bipartite graph that connects all possible pairs of nodes across two different groups, with $m$ and $n$ nodes in them, while connecting no edges at all between nodes in the same group. Therefore, $K_{1,3}$ is just a star with 3 edges. This graph has one center and it connects to three other nodes, with no other connections.

In the following section, we will use the line graph of the complex PPI network for RMC graph alignment. 

\section{Results}
\label{sec:results}
\subsection{Applying RMC to a Torus}
\begin{figure}
\centering
\includegraphics[width=0.6\textwidth]{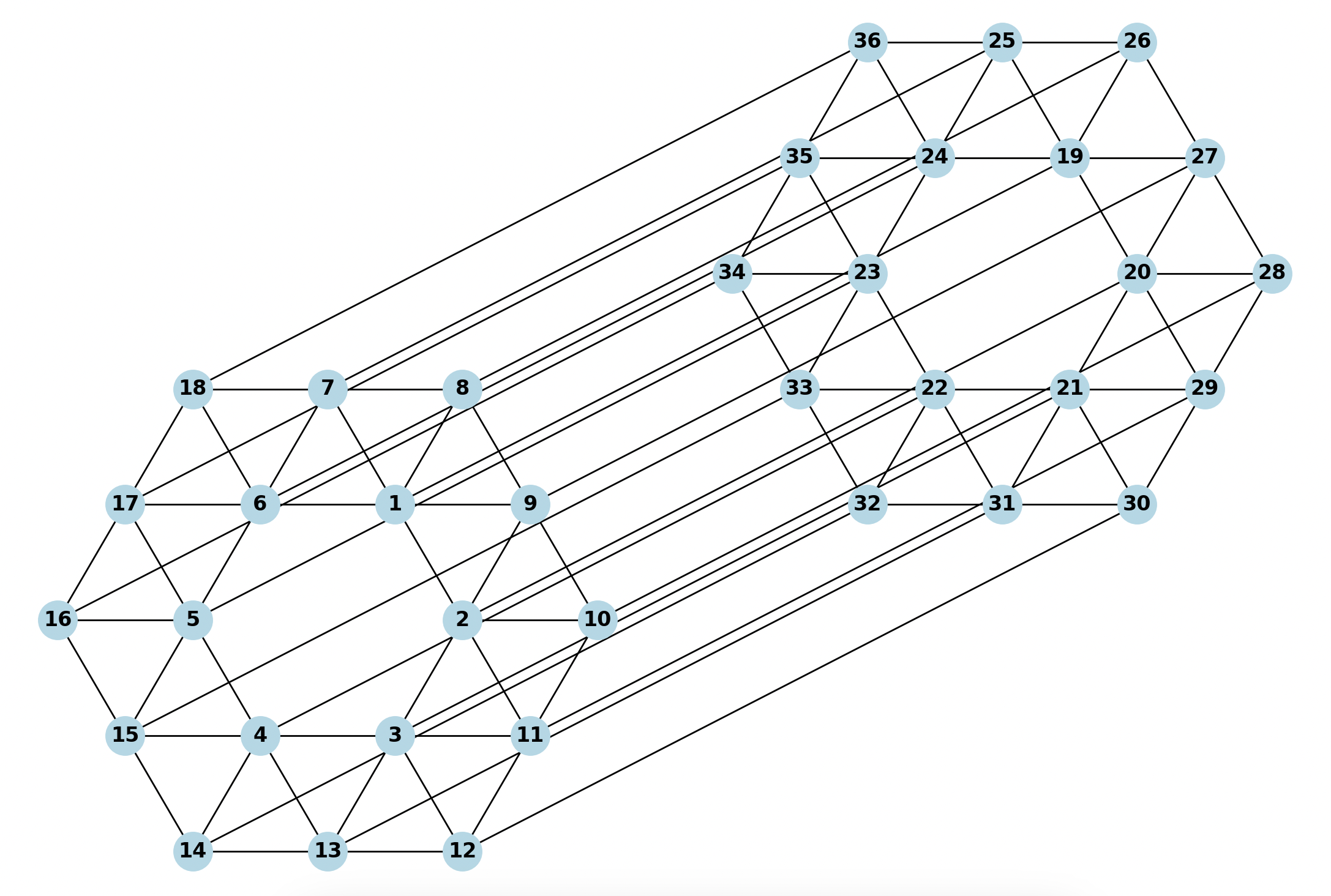}
  \caption{3D triangular tiling of torus.}
  \label{fig:3DTriangTorus}
\end{figure}

\begin{figure}
\centering
\includegraphics[width=0.45\textwidth]{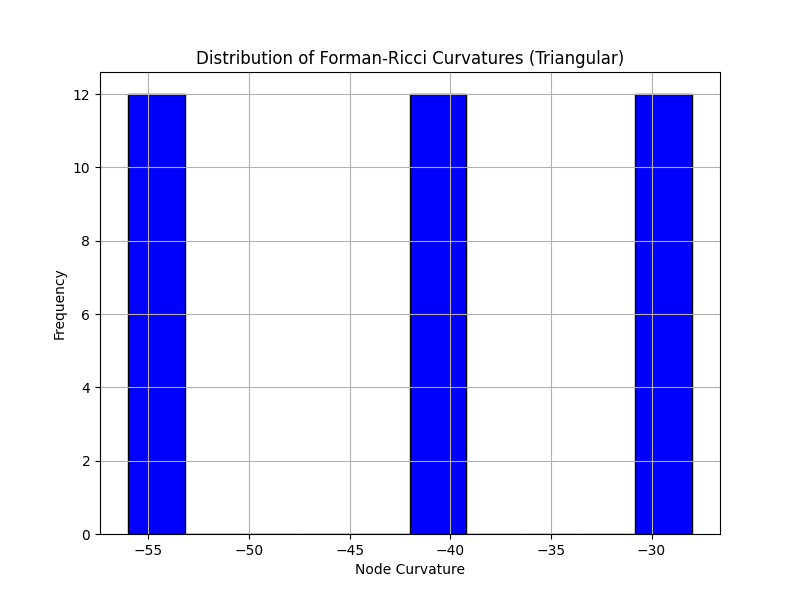}
  \caption{Distribution of curvature in triangle-tiled torus.}
  \label{fig:3DTriangDist}
\end{figure}

\paragraph{Triangular Tiling} We examine the effects of applying RMC to a tiled torus created by first creating a two-dimensional triangular tiling as described in Section~\ref{sec:2DTiling} and then lifting it (see Figure~\ref{fig:3DTriangTorus}). 
If we examine the node curvatures of this graph, we find the following spectrum of distribution (see Figure~\ref{fig:3DTriangDist}). There are three types of nodes in the torus, and we find that each of their Ricci matrix row vector will be different, which will help us differentiate the three types of nodes at different locations of the graph. Let us call the nodes with smallest curvature $A$, medium curvature nodes $B$, and the rest $C$. $A$ is closest to the center (bordering the hole of the torus), $B$ and $C$ are on the outer brink of the ``doughnut'' shape. Because of the discretized implementation, $B$ and $C$ become different, even though in the continuous case they should be the same. But we see that $|\textbf{Ric}(B) - \textbf{Ric}(C)| < |\textbf{Ric}(A) - \textbf{Ric}(B)|$. Then the Ricci matrix row vectors for nodes $A,B,C$ are $[\textbf{Ric}(A), \textbf{Ric}(A),\textbf{Ric}(A),\textbf{Ric}(B),\textbf{Ric}(B),\textbf{Ric}(C) ]$, $[\textbf{Ric}(A), \textbf{Ric}(A),\textbf{Ric}(B),\textbf{Ric}(C),\textbf{Ric}(C),0]$, and $[\textbf{Ric}(A), \textbf{Ric}(B),\textbf{Ric}(B),\textbf{Ric}(C),0,0]$, respectively. If we apply RMC to two tori of this type, we will be able to completely align the hole of the torus to the hole on another torus with the same shape, since the $A$ nodes are located at and only located at the ``boundaries'' of the hole (see for example, nodes labeled 1 to 6 in Figure~\ref{fig:3DTriangTorus}). We do not elaborate on the square tiling and mixed tiling cases as the methodology to approach them is identical. We will get distinct row vectors for nodes at different locations (by ``different'' we mean geometrically non-isomorphic). 

\subsection{Applying RMC to Line Graph of Protein-Protein Interaction Network}
As mentioned in our introduction section, the protein-protein interaction (PPI) network is a common type of network used to test alignment methods. We used a modified version of the dataset in SNAP: ``Feature Learning in Multi-Layer Networks'' \citep{Zitnik2017}, provided as ``combined\_ppi.graphml'' in \citet{Wang2025}. We transform this PPI network to its line graph form (see Section~\ref{sec:linegraph}) using the \texttt{line\_graph} function from \texttt{networkx}. Since the line graph transformation is expensive, we first take a 1000-node graph from the original PPI network through random walk, then transform that sampled graph into a line graph. Afterwards, we sample two 500-node subgraphs for alignment using RMC. This is done by doing random walk on the 1000-node graph just obtained, and then we create a second graph for alignment through random deletion of edges of the first 500-node graph at a probability of $p = 0.01$. We use the same random walk method and alignment set up as in \citet{Wang2025}. We run the same experiment for ten times (see Table~\ref{sample-table}). Furthermore, we note that the CPU model used was Apple M1 with 16GB of memory. In particular, we implemented the experiments using Python 3.9.6 on Pycharm. Running the entire process, from random walk, to creating the line graph, and the graph alignment took around twenty hours. Graph alignment for ten times takes less than an hour. If we refer to Table 1 in \citet{Wang2025} for baseline results, we see that RMC is more accurate than most except for DMC and is at a similar level as REGAL. While the line graph generation is lengthy, the time complexity of RMC itself is on the same order as DMC, at around $O(n^{2})-O(n^{3})$.

\begin{table}
  \caption{RMC on Line Graph of PPI Network}
  \label{sample-table}
  \centering
  \begin{tabular}{lll}
    \toprule
    Round     & Absolute Node Count     & Percentage \\
    \midrule
    1 & 416  & 83.2\%    \\
    2 & 445  & 89.0\%    \\
    3 & 464  & 92.8\%    \\
    4 & 448  & 89.6\%    \\
    5 & 452  & 90.4\%    \\
    6 & 476  & 95.2\%    \\
    7 & 438  & 87.6\%    \\
    8 & 462  & 92.4\%    \\
    9 & 432  & 86.4\%    \\
    10 & 460  & 92.0\%    \\
    \bottomrule
  \end{tabular}
\end{table}

\section{Conclusion and Future Work}

In this paper, we introduced a new setting for graph alignment, utilizing the concept of curvature. Previous work proposed a degree-based alignment method, and we have modified it, after gaining sufficient theoretical motivations, to use the Forman-Ricci curvature as a main feature of nodes. We also discussed possible smooth surface discretization methods that utilize regular shapes as units. This discussion was necessary to fit graph alignment into the differential geometric view. In the search for graphs that are suitably described by curvatures, we not only examined the discretization of a torus, but also proposed using line graphs of complex networks for RMC alignment. Results showed that the RMC has potential to be more widely studied and used. Future work can experiment with RMC on a wider variety networks and possibly derive theoretical guarantees for performance, which is possible since there are initial theoretical guarantees for DMC. Moreover, we suspect the Curvature-Laplacian equation to be related to the Euler characteristic.

% \begin{ack}
% Use unnumbered first level headings for the acknowledgments. All acknowledgments

% \end{ack}

\bibliographystyle{plainnat}
\bibliography{ricci}

\appendix
\section{Torus Creation}
The code in this section shows how we manually constructed our torus and plotted the curvature distribution. 
\begin{lstlisting}
import networkx as nx
import matplotlib.pyplot as plt

# we manually create a torus that is triangulated

# node positions
positions = {
    1: (1, 1), 2: (2, 0),
    3: (1, -1), 4: (-1, -1),
    5: (-2, 0), 6: (-1, 1),
    7: (0, 2), 8: (2, 2),
    9: (3, 1), 10: (4, 0),
    11: (3, -1), 12: (2, -2),
    13: (0, -2), 14: (-2, -2),
    15: (-3, -1), 16: (-4, 0),
    17: (-3, 1), 18: (-2, 2)
}

# edges
edges = [
    (1, 2), (1, 6), (1, 7), (1, 8), (1, 9),
    (2, 9), (2, 10), (2, 11), (2, 3),
    (3, 11), (3, 12), (3, 13), (3, 4),
    (4, 13), (4, 14), (4, 15), (4, 5),
    (5, 15), (5, 16), (5, 17), (5, 6),
    (6, 17), (6, 18), (6, 7), (7, 8),
    (7, 18), (8, 9), (9, 10), (10, 11),
    (11, 12), (12, 13), (13, 14), (14, 15),
    (15, 16), (16, 17), (17, 18)
]

# original graph
G = nx.Graph()
G.add_nodes_from(positions.keys())
G.add_edges_from(edges)

# create a copy of the graph and shift its position
offset_x, offset_y = 10, 3
new_positions = {}
mapping = {}
for old_id in positions:
    new_id = old_id + 18
    mapping[old_id] = new_id
    x, y = positions[old_id]
    new_positions[new_id] = (x + offset_x, y + offset_y)

# add new nodes and edges
G.add_nodes_from(new_positions.keys())
G.add_edges_from([(mapping[u], mapping[v]) for u, v in edges])

# correspond nodes on two copies
G.add_edges_from([(u, mapping[u]) for u in positions])

# combine all positions
all_positions = {**positions, **new_positions}

# draw the graph
plt.figure(figsize=(12, 8))
nx.draw(G, pos=all_positions, with_labels=True, node_color='lightblue', node_size=500, font_weight='bold')
plt.title("Graph of Torus")
plt.show()

## plotting curvature distribution
def compute_edge_forman_curvature(G):
    return {(u, v): 2 - G.degree[u] - G.degree[v] for u, v in G.edges()}

def compute_node_forman_curvature(G, edge_curvatures):
    node_curvature = {v: 0 for v in G.nodes()}
    for (u, v), curvature in edge_curvatures.items():
        node_curvature[u] += curvature
        node_curvature[v] += curvature  # symmetric edge
    return node_curvature

# compute the two types of curvatures
edge_curvatures = compute_edge_forman_curvature(G)
node_curvatures = compute_node_forman_curvature(G, edge_curvatures)

# plot distribution
curv_values = list(node_curvatures.values())

plt.figure(figsize=(8, 6))
plt.hist(curv_values, bins=10, color='blue', edgecolor='black')
plt.title("Distribution of Forman-Ricci Curvatures (Triangular)")
plt.xlabel("Node Curvature")
plt.ylabel("Frequency")
plt.grid(True)
plt.show()
\end{lstlisting}

\section{Applying RMC to Line Graph of PPI}
The code in this section involves construction of PPI line graph and aligning with RMC. One will need to download the file ``combined\_ppi.graphml'' and manually fill in the path to this file on local machine that is provided with this paper, originally from \citet{Wang2025}. The SNAP dataset has a BSD license and the modified dataset has a license of CC-BY 4.0.
\begin{lstlisting}
import networkx as nx
import numpy as np
from scipy.optimize import linear_sum_assignment
from itertools import permutations
import random

# random walk, get line graph, sample two subgraphs

# Function to load G_{r} from a GraphML file
def load_graph_from_graphml(graphml_path):
    graph = nx.read_graphml(graphml_path)
    graph = random_walk(graph, 1000)
    graph = nx.line_graph(graph)

    mapping = {node: idx for idx, node in enumerate(graph.nodes())}
    G_relabel = nx.relabel_nodes(graph, mapping)
    return G_relabel


# Random Walk sampling method
def random_walk(mygraph, subgraph_size, max_iter=100):
    progress = 0
    current_node = np.random.choice(list(mygraph.nodes()))
    network_sub_nodes = [current_node]

    iterations = 0
    while len(network_sub_nodes) < subgraph_size:
        neighbors = list(mygraph.neighbors(current_node))

        if neighbors:
            next_node = np.random.choice(neighbors)
        else:
            next_node = np.random.choice(list(mygraph.nodes()))

        # check if the next node is already in the subgraph
        if next_node not in network_sub_nodes:
            network_sub_nodes.append(next_node)
            progress += 1
            iterations = 0  # reset iteration counter since progress was made
        else:
            iterations += 1  # increment if no new node was added

        # If stuck, reset to a random new node after max_iter iterations.
        if iterations >= max_iter:
            potential_nodes = set(mygraph.nodes()) - set(network_sub_nodes)
            if potential_nodes:
                next_node = np.random.choice(list(potential_nodes))
            else:
                print("No more new nodes to reset to; exiting.")
                break
            iterations = 0  # reset iteration count after switching nodes

        # Update the current node
        current_node = next_node

    network_sub = mygraph.subgraph(network_sub_nodes).copy()
    # print("Random walk completed.")
    return network_sub


# randomly delete edges with specified probability
def delete_edges_randomly(graph, deletion_probability):
    modified_graph = graph.copy()
    edges_to_remove = [edge for edge in graph.edges if random.random() < deletion_probability]
    modified_graph.remove_edges_from(edges_to_remove)
    return modified_graph


def ricci_curvature(G, node):
    neighbors = list(G.neighbors(node))
    if neighbors:  # Check if the node has neighbors
        curvature = np.sum([(2-G.degree(neighbor)-G.degree(node)) for neighbor in neighbors])
    else:
        curvature = 0
    return curvature


# Create degree matrix for alignment
def create_neighbor_degree_matrix(graph, max_degree):
    num_nodes = len(graph.nodes())
    matrix = np.zeros((num_nodes, max_degree))

    for i, node in enumerate(graph.nodes()):
        # use Ricci curvature instead of neighbor degree sorted
        neighbors = list(graph.neighbors(node))
        neighbor_curvatures = sorted([ricci_curvature(graph, neighbor) for neighbor in neighbors], reverse=True)

        matrix[i, :len(neighbor_curvatures)] = neighbor_curvatures

    return matrix, list(graph.nodes())


# Align two matrices for graph alignment (Hungarian algorithm)
def align_matrices(matrix1, matrix2):
    cost_matrix = np.zeros((matrix1.shape[0], matrix2.shape[0]))

    for i in range(matrix1.shape[0]):
        for j in range(matrix2.shape[0]):
            cost_matrix[i, j] = np.linalg.norm(matrix1[i] - matrix2[j])

    row_ind, col_ind = linear_sum_assignment(cost_matrix)
    return row_ind, col_ind


# Perform graph alignment
def graph_alignment(graph1, graph2):
    max_degree = max(
        max([len(list(graph1.neighbors(node))) for node in graph1.nodes()]),
        max([len(list(graph2.neighbors(node))) for node in graph2.nodes()])
    )
    matrix1, nodes1 = create_neighbor_degree_matrix(graph1, max_degree)
    matrix2, nodes2 = create_neighbor_degree_matrix(graph2, max_degree)

    row_ind, col_ind = align_matrices(matrix1, matrix2)
    node_mapping = {nodes1[i]: nodes2[j] for i, j in zip(row_ind, col_ind)}
    return node_mapping


# compute the signature vector for node equivalence (use this if one wants to count geometrically)
def compute_signature_vector(graph, node):
    neighbors = list(graph.neighbors(node))
    signature_vector = []

    for neighbor in neighbors:
        neighbor_degrees = [graph.degree(n) for n in graph.neighbors(neighbor)]
        signature_vector.append(sorted(neighbor_degrees))

    return signature_vector


# check if two nodes are equivalent by comparing their signature vectors
def are_nodes_equivalent(graph, node1, node2):
    if graph.degree(node1) in [1, 2, 3] and graph.degree(node2) in [1, 2, 3]:
        signature1 = compute_signature_vector(graph, node1)
        signature2 = compute_signature_vector(graph, node2)
        for perm in permutations(signature1):
            if list(perm) == signature2:
                return True
            else:
                return False
    else:
        return False


# Calculate correct predictions based on alignment
def calculate_correct_predictions(alignment, graph):
    correct_prediction_count = 0
    for node1 in alignment:
        predicted_node2 = alignment[node1]
        if predicted_node2 == node1:
                # use the following line if we count geometrically
                # for now we count by node id
                #or are_nodes_equivalent(graph, node1, predicted_node2):
            correct_prediction_count += 1
    return correct_prediction_count


# full alignment experiment
def run_alignment_experiment(graph, target_node_count, deletion_probability):
    sampled_subgraph = random_walk(graph, target_node_count)
    print("First subgraph created.")
    print("G1 node count: ", len(sampled_subgraph.nodes()))
    print("G1 edge count: ", len(sampled_subgraph.edges()))

    # randomly delete edges in G_{1} to get G_{2}
    final_subgraph = delete_edges_randomly(sampled_subgraph, deletion_probability)
    print("Final G2 subgraph created.")
    print("G2 node count: ", len(final_subgraph.nodes()))
    print("G2 edge count: ", len(final_subgraph.edges()))

    # apply RMC
    alignment = graph_alignment(sampled_subgraph, final_subgraph)

    # the number of correct predictions
    correct_predictions = calculate_correct_predictions(alignment, sampled_subgraph)
    return correct_predictions


# Main execution
graphml_path = "/Path/to/combined_ppi.graphml"
main_graph = load_graph_from_graphml(graphml_path)

target_node_count = 500
deletion_probability = 0.01

round = 0
for i in range(10):
    round += 1
    print("Round ", round)
    correct_predictions = run_alignment_experiment(main_graph, target_node_count, deletion_probability)
    print("Correct predictions:", correct_predictions)
\end{lstlisting}

\end{document}